# Interpretation of the Area Under the ROC Curve for Risk Prediction Models

Ralph H. Stern

Division of Cardiovascular Medicine

Department of Internal Medicine

University of Michigan

Ann Arbor, Michigan

stern@umich.edu

## Abstract

The area under the curve (AUC) of the receiver operating characteristics curve (ROC) evaluates the separation between patients and nonpatients or discrimination. For risk prediction models these risk distributions can be derived from the population risk distribution so are not independent as in diagnosis. A ROC curve AUC formula based on the underlying population risk distribution clarifies how discrimination is defined mathematically and that generation of the equivalent c-statistic effects a Monte Carlo integration of the formula. For a selection of continuous risk distributions, exact analytic formulas or numerical results for the ROC curve AUC and overlap measure are presented and demonstrate a linear or near-linear dependence on their standard deviation. The ROC curve AUC is also shown to be highly dependent on the mean population risk, a distinction from the independence from disease prevalence for diagnostic tests. The converse of discrimination, overlap, has been quantified by the overlap measure, which appears to provide equivalent information. As achieving wider population risk distributions is the goal of risk prediction modeling for clinical risk stratification, interpreting the ROC curve AUC as a measure of dispersion, rather than discrimination, when comparing risk prediction models may be more relevant.

The receiver operating characteristics (ROC) curve was introduced in medicine [1] to evaluate diagnostic tests. The area under the curve (AUC) is used to measure the separation of patients and nonpatients, i.e. discrimination. Subsequently, the ROC curve AUC has been used to evaluate discrimination of risk prediction models. However the differences between these applications has generally been ignored. Cook [2] pointed out that they serve different purposes: "accurately identifying an existing, but unknown, disease state" versus "prediction of future risk in a normal population" that can be used for "risk stratification, or for assigning levels of risk...which may then form the basis of treatment decisions." [2, p. 17] Risk prediction models applied to a population generate population risk distributions that in turn generate patient and nonpatient risk distributions that are thus not independent, so the differences between diagnostic and prognostic models extend beyond purpose. A wider distribution of risk in the population when presented graphically has previously been identified as a sign of better risk stratification [3] Here it is shown using selected continuous distributions that the ROC curve AUC for risk prediction models is directly related to this widening. Thus the ROC curve AUC can be interpreted as a measure of dispersion of the population risk distribution when comparing risk prediction models, not just a measure of discrimination of the patient and nonpatient risk distributions.

METHODS

Derivations, calculations, and graphs were produced with computer algebra systems (Maple 2017, Maplesoft, Waterloo, Ontario and/or Mathematica, Wolfram Research, Champaign, Illinois)

If the distribution of risk in the population is f($r$), the prevalence of patients and nonpatients at any level of risk is $r$ f($r$) and (1-$r$) f($r$), respectively The patient and nonpatient risk distribution curves or probability distribution functions (pdfs) are $r$ f($r$)/$r_{mean}$ and (1-$r$) f($r$)/(1-$r_{mean}$), respectively, where $r_{mean}$ is the mean risk for the population. The true positive fraction is the fraction of patients above a threshold risk (t), while the false positive fraction is the fraction of nonpatients above a threshold risk.

$$TP = \int_t^1 \frac{r\, f(r)}{r_{mean}}\, dr$$

$$FP = \int_t^1 \frac{(1-r)\, f(r)}{1-r_{mean}}\, dr$$

The ROC curve presents the true positive fraction (sensitivity) versus the false positive fraction (1-specificity) for varying threshold risks. Since both true positives and false positives depend on the threshold risk, the ROC curve is a parametric curve. The formula for the area under the curve for a parametric curve gives [4]:

$$AUC = \int TP\, dFP = \int_0^1 TP(r)\, FP'(r)\, dr = \int_0^1 \left(\frac{(1-r)\, f(r)}{1-r_{mean}}\right)\left(\int_r^1 \frac{x\, f(x)}{r_{mean}}\, dx\right) dr$$

where $x$ is a dummy variable to avoid using $r$ both in the integration limit and the integrand.

The outer integral integrand is the nonpatient risk distribution, while the inner integral integrand is the patient risk distribution. The area under the nonpatient risk distribution is one, so the outer integral is one. As the formula calculates for every level of risk the product of the frequency of nonpatients and the fraction of the patient distribution above that level of risk, it is a mathematical measure of the overlap of the two distributions. If the distributions did not overlap, the value of the inner integral would be one for all risk values and the AUC would be one. If the distributions were superimposable, the value of the inner integral would decrease proportionally from one to 0 and the AUC would be 1/2.

The c-statistic [5] is obtained by repetitively selecting a patient and nonpatient and determining the fraction of pairs for which the risk prediction model correctly ordered the pair. In the limit, the results could be represented with integrals based on the patient and nonpatient risk distributions. As the formula calculates for every level of risk the product of the frequency of nonpatients and the fraction of patients above that level of risk (i.e. correctly ordered), it is equivalent to the c-statistic, demonstrating that generation of the c-statistic effects a Monte Carlo integration of the above AUC formula.

The converse of discrimination, overlap, is infrequently considered. It can be quantified by the following expression, called the overlap measure (OM) in a recent publication [6], which was described as early as 1937. [7] For the two risk distribution curves (nonpatient and patient):

$$OM = \int_0^1 min[nonpatient(r), patient(r)]\, dr = \int_0^1 min\left[\left(\frac{r\, f(r)}{r_{mean}}\right), \left(\frac{(1-r)\, f(r)}{1-r_{mean}}\right) dr\right]$$

The OM ranges from one for two completely overlapping probability distribution curves to 0 for two completely separated probability distribution curves. The patient and nonpatient risk distribution curves intersect when: $r\, f(r)/r_{mean} = (1-r)\, f(r)/(1-r_{mean})$ which occurs at $r_{mean.}$ Since the patient risk distribution curve lies below the nonpatient risk distribution curve before the intersection and the nonpatient risk distribution curve lies below the patient risk distribution curve after the intersection:

$$OM = \int_0^{r_{mean}} \frac{(r)\, f(r)}{r_{mean}}\, dr + \int_{r_{mean}}^1 \frac{(1-r)\, f(r)}{1-r_{mean}}\, dr$$

The Table presents the formulas for uniform, half-sine, lognormal, and beta distributions for the population pdf's; their means; and their standard deviations. Analytic expressions for the ROC curve AUC and OM are presented for the uniform and half sine distributions.

The parameters for a lognormal distribution are $m$ (often $\mu$) and $s$ (often $\sigma$), where $m$ is the mean and $s$ is the standard deviation (SD) of the normal distribution obtained by taking the log of the lognormal curve (as opposed to the mean and standard deviation of the lognormal distribution). The mean of a lognormal distribution, $\exp(m+s^2/2)$, depends on both parameters, as does the standard deviation of the lognormal distribution. To explore dependence of AUC on SD for a given mean, the formula for the AUC was reparametrized using $r_{mean}$.

Lognormal population risk distributions can extend well past one resulting in a departure from linearity with more disperse population risk distribution curves, followed by a plateau and descent. This departure becomes apparent when the 99th percentile of the patient risk distribution equals one, so the corresponding standard deviation was adopted as an upper limit for evaluation.

For the uniform and half-sine distributions, the graphs extend to the highest possible standard deviation for the presented means, while the range of standard deviations for the beta distribution was selected to provide similar AUCs and OMs.

The analytic formulas for uniform and half-sine distributions were used. Numerical methods for lognormal and beta distributions were used

RESULTS

Figure 1 presents AUC and OM vs standard deviation for each of the distributions with mean risks of 0.01, 0.05, 0.1, and 0.2. The plots of the ROC curve AUC and OM versus standard deviation are linear or near linear, with the relationship dependent on the mean risk of the distribution.

The AUC and OM appear to provide equivalent information. For the uniform and half-sine distributions, equating the common terms in the AUC and OM formulas allows expressing the AUC in terms of the OM.

uniform distribution: AUC=(7/6)-(2/3) x OM

half-sine distribution: AUC=((3-OM) $\pi - 4)/(4\pi - 8)$

These relationships are valid across possible distributions (i.e. up to a range of risks between 0 and one). When OM=1 is inserted, the correct AUC of 1/2 is obtained. But an OM of 0 is not possible for a risk distribution curve as it would require all patients to have a risk of one and all nonpatients a risk of 0. From these equations, the OM and AUC lines intersect when both reach a value of 0.7. Graphically, it appears the same is also the case for the lognormal and beta distributions.

DISCUSSION

Simulations may provide clearer insight into underlying principles than real data. And mathematically simple distributions may permit analytic solution. The uniform distribution is not a realistic risk distribution, but the half-sine, and especially the lognormal [8] and beta distribution are.

For all of the distributions and each of the mean risks (0.01,0.05, 0.1, and 0.2), at zero standard deviation all AUC's are 0.5 and increase linearly with SD. In contrast, at zero standard deviation, all OM's are one and decrease linearly with SD. For a given distribution, the slopes of the lines for the different means differ. Overall the AUC and OM appear to provide equivalent information, although the underlying calculations appear distinct.

In a different context the mathematical relationship between the OM (which was called $\Delta_1$) and the AUC/c-statistic (which was called $\Delta_2$) for two independent probability distribution functions (test and control) that were normally distributed with the same standard deviation ($\sigma$), but different means ($\mu^T > \mu^C$) with $\delta=(\mu^T-\mu^C)/\sigma$ has been described [7]:

OM=$\Delta_1$=200 $\Phi(-\delta/2)$

$AUC=\Delta_2=1-\Phi(-\delta/\sqrt{2})= \Phi(\delta/\sqrt{2})$

where $OM=\Delta_1$ is in percent and $\Phi(x)$ is the standard normal cumulative distribution function. However these binormal equations do not apply to the probability distribution functions for patients and nonpatients generated from a single population probability distribution function.

The ROC curve AUC for diagnostic tests does not depend on disease prevalence, so the dependence on mean risk is a major difference. The sensitivity and specificity of a diagnostic test are independent, but the risk distributions of patients and nonpatients are not and the location of the population risk distribution markedly influences their overlap. Two uniform population risk distributions with different mean risks, but the same standard deviation, are depicted in Figure 2: U(0,0.02) and U(0.49,0.51). The uniform distribution with a mean risk of 0.01 generates very different patient and nonpatient risk distributions with a ROC curve AUC of 0.668, whereas the uniform distribution with a mean risk of 0.5 generates almost identical patient and nonpatient risk distributions with a ROC curve AUC of 0.507.

Janssens and Martens recently explored [9] the interpretation of the ROC curve as an alternative way of presenting the risk distributions of diseased (patients) and non-diseased (nonpatients) individuals. Their analysis refuted criticisms of the ROC curve AUC as clinically irrelevant and lacking an intuitive interpretation. The simulations presented here confirm and extend their conclusions.

Repetitively sampling patients and nonpatients and determining the fraction of comparisons in which a model correctly rank orders their risk is shown to represent a Monte Carlo integration of the formula for the ROC curve AUC, rather than a bizarre, clinically irrelevant experiment.

Turning to the underlying population risk distribution curve, as opposed to the derivative patient and nonpatient risk distribution curves, the ROC curve AUC for a given population is shown to be a function of the underlying shape of the risk distribution, its mean, and its dispersion. The ROC curve AUC dependence on the mean risk contrasts with the its independence of disease prevalence for diagnostic tests, where sensitivity and specificity are independent and the distributions of test results in patients and nonpatients are unlinked.

It is intuitively clear that the more disperse the parent population risk distribution, the greater the separation of the derivative patient and nonpatient risk distributions. When diagnostic testing, the resolution of patients and nonpatients is the goal, i.e. separating qualitatively distinct patient populations (distinguishing black and white). This permits treating patients, while avoiding the treatment of nonpatients. Thus the ROC curve AUC is a natural metric. The analogous discriminative ability of risk prediction models has been emphasized. On the other hand, in prediction, maximal dispersion of the population risk distribution is the goal, i.e. separating quantitatively distinct patient populations (distinguishing shades of grey). The wider the variation in risk when the risk distribution in the population is visualized, the better the risk stratification [3] with potential clinical benefit. Thus reconsidering the ROC curve AUC to be a measure of this dispersion when comparing risk prediction models is intuitively appealing and supported by its linear dependence on the standard deviation.

In the past, concerns about the ROC curve AUC have been raised when clearly established risk factors added to risk prediction models did not increase the ROC curve AUC [2]. However if the population risk distribution curve is not broader, which is what the ROC curve AUC measures, there is no potential clinical benefit from the addition.

## FIGURE LEGENDS

Figure 1

Area under the curve (AUC) and overlap measure (OM) vs standard deviation (SD) for population risk distributions with mean risks of 0.01, 0.05, 0.1, and 0.2. (thicker lines correspond to increasing mean risks.)

Figure 2

Patient and nonpatient risk distributions for uniform population risk distributions with mean risks of 0.01 and 0.5 (thick lines patients, thin lines nonpatients)

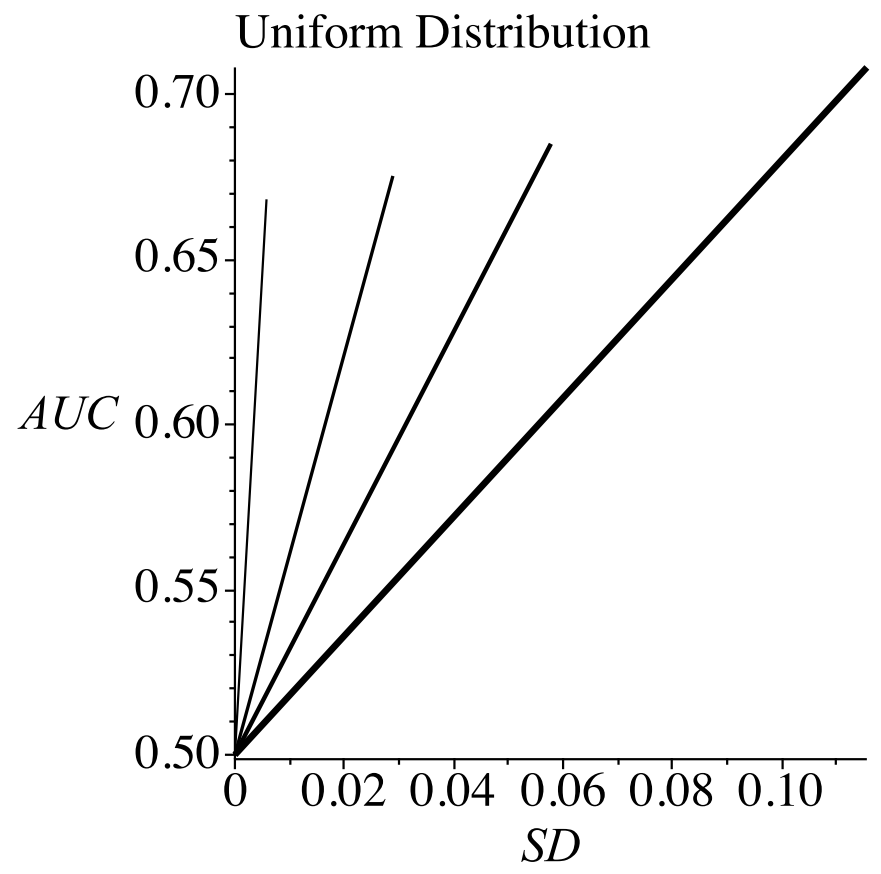

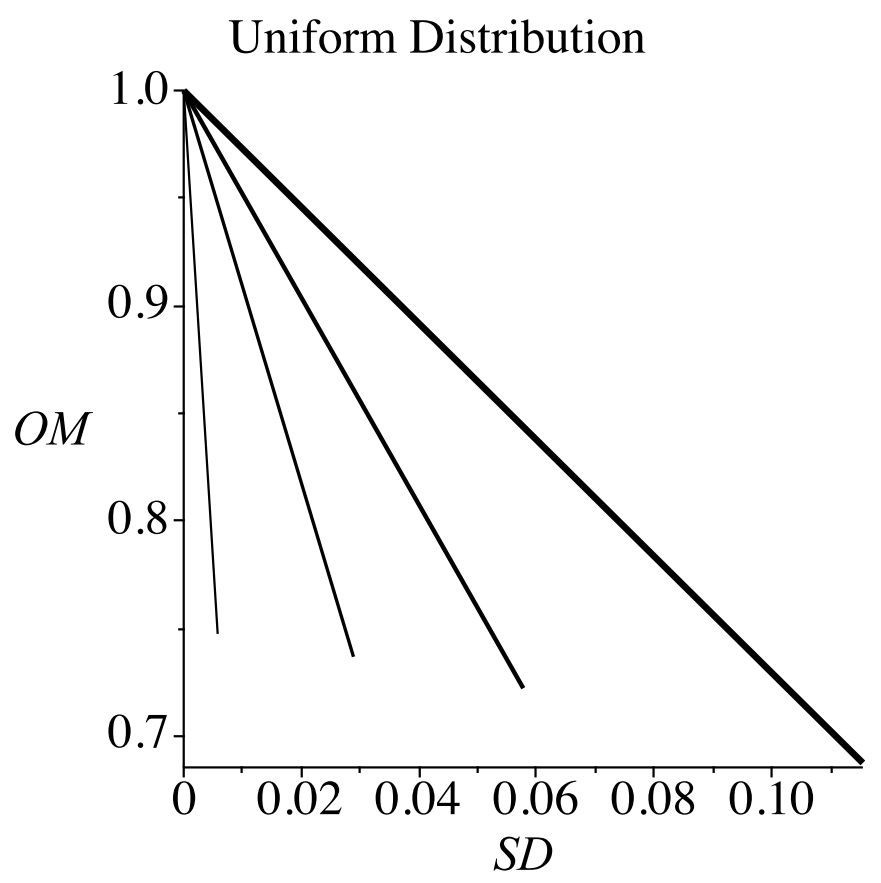

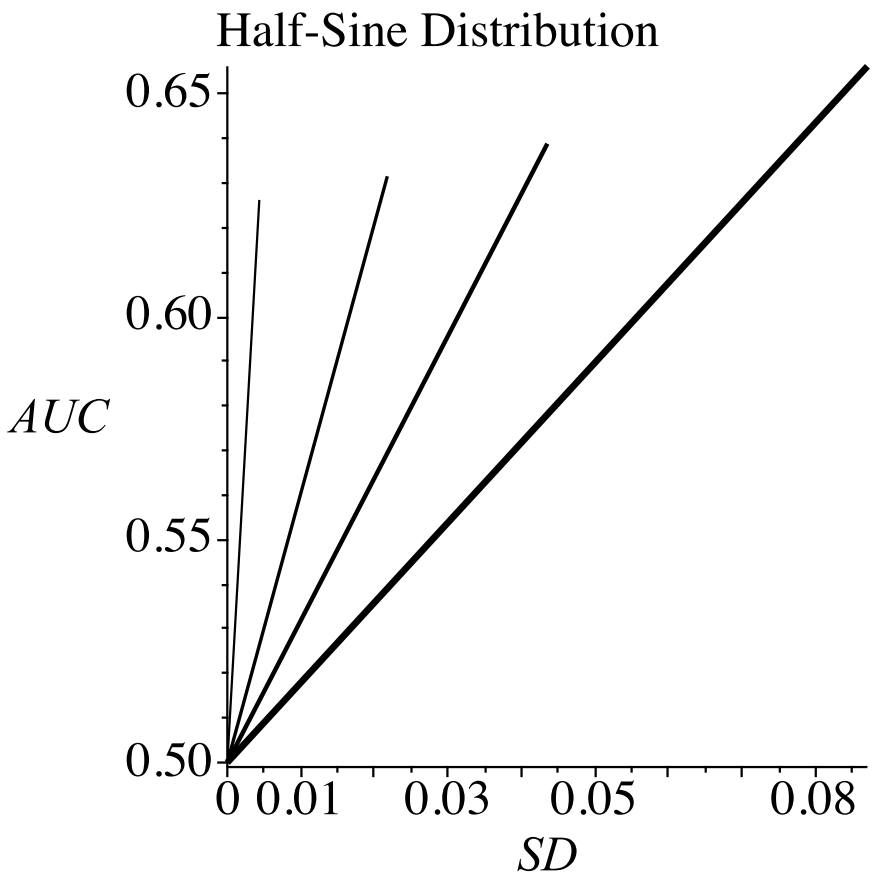

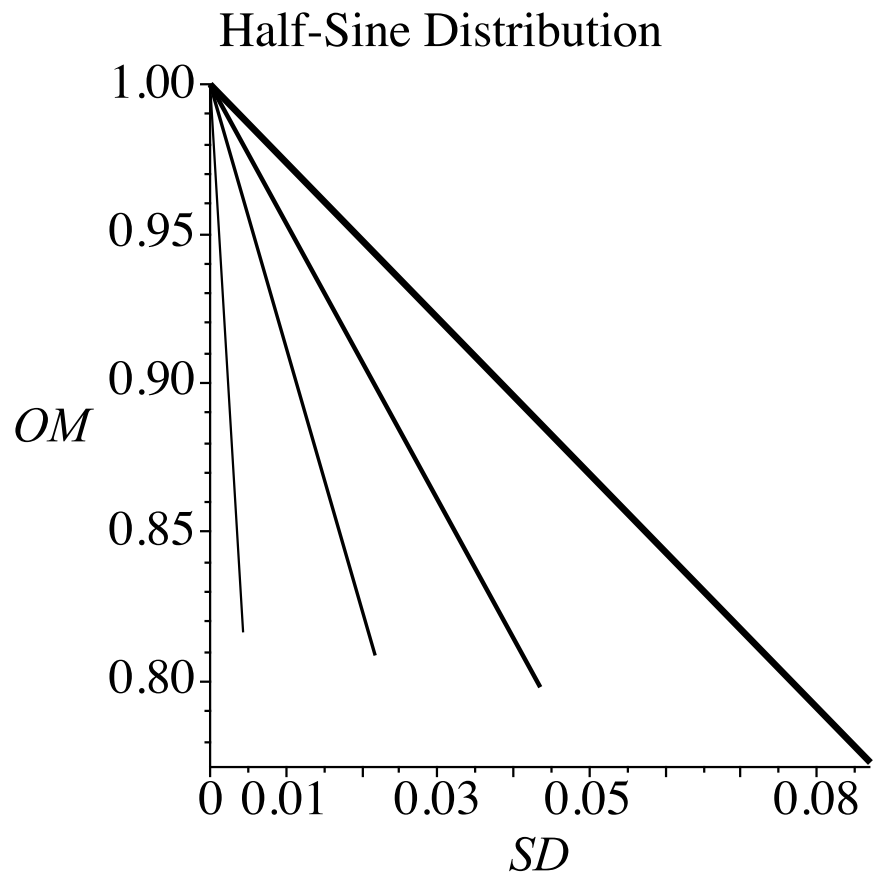

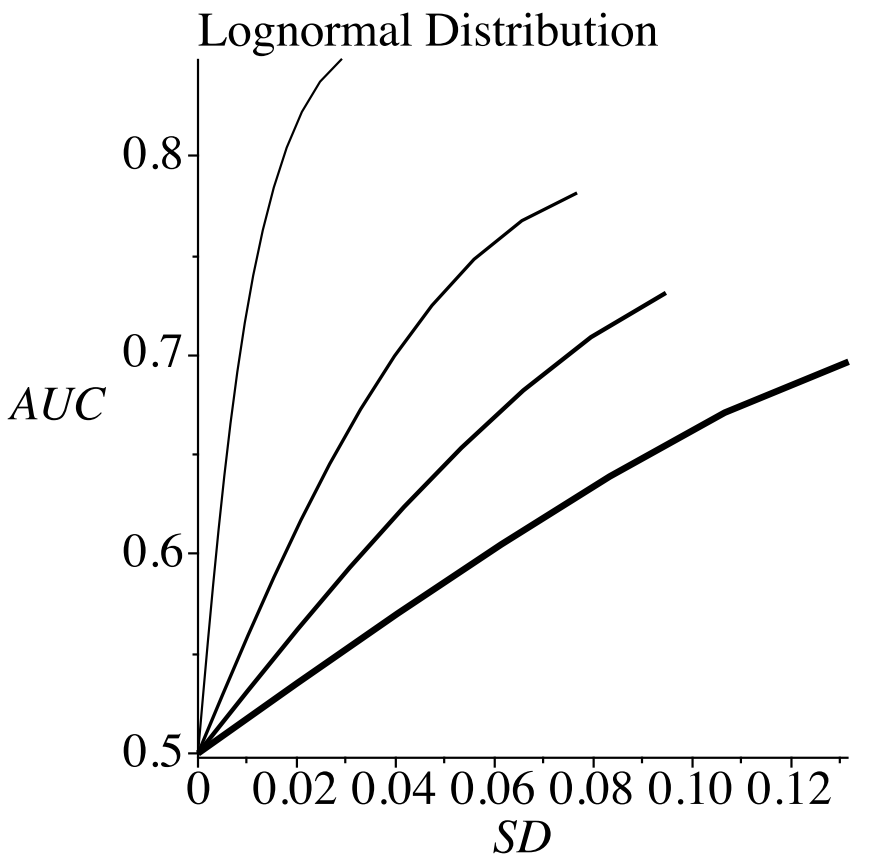

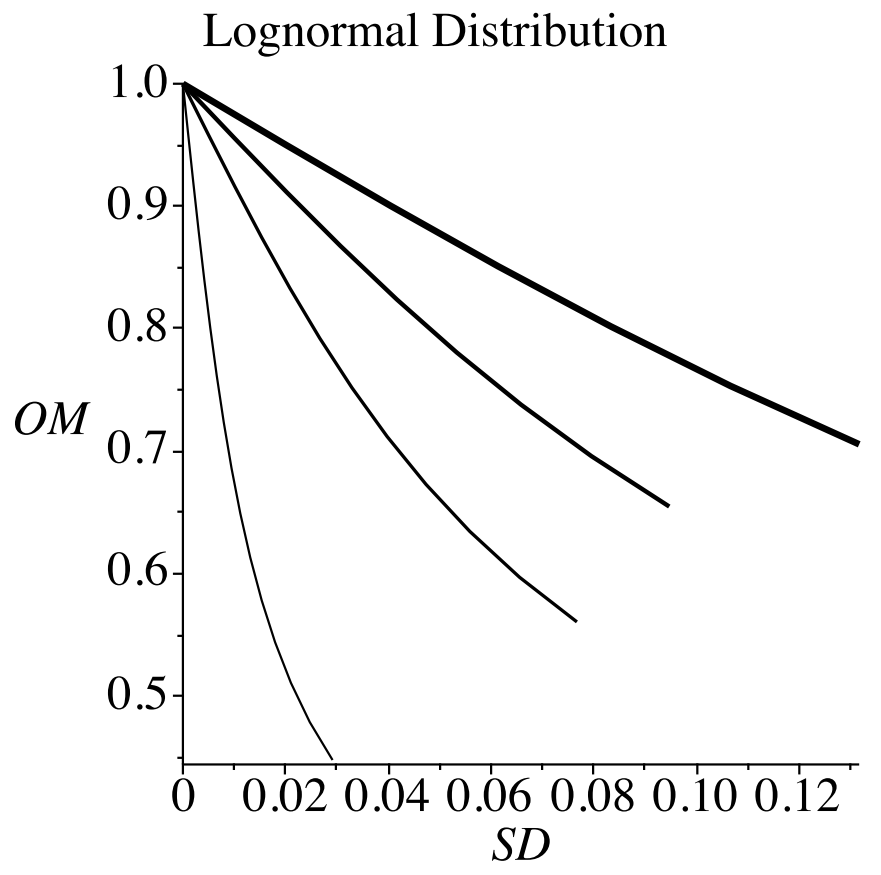

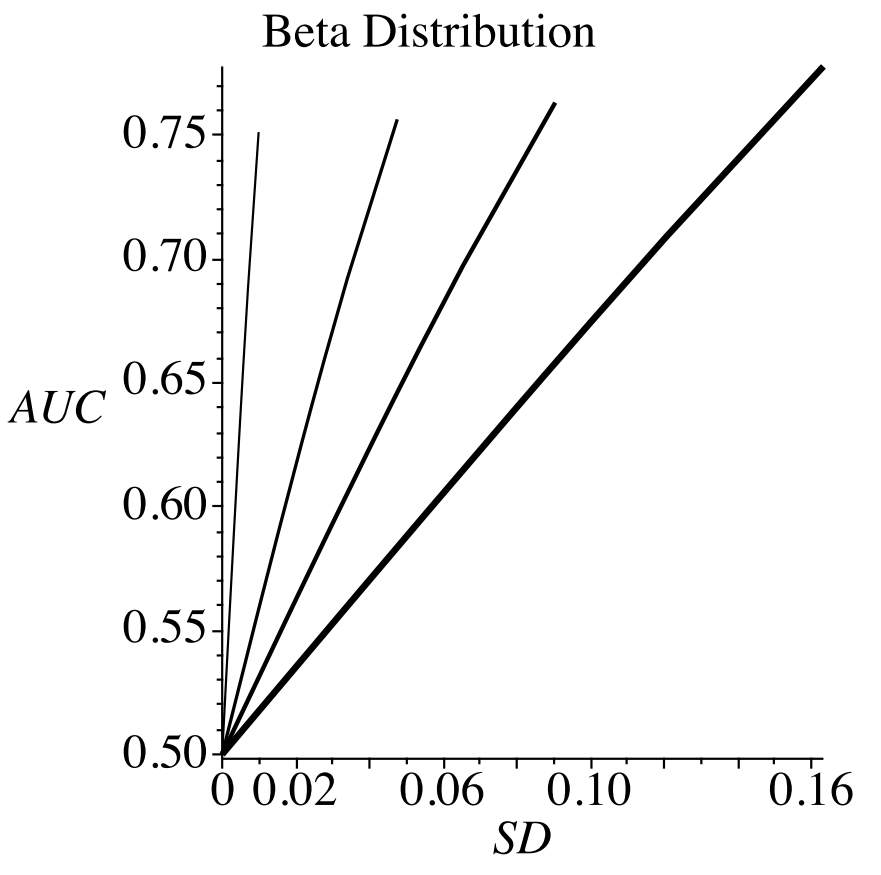

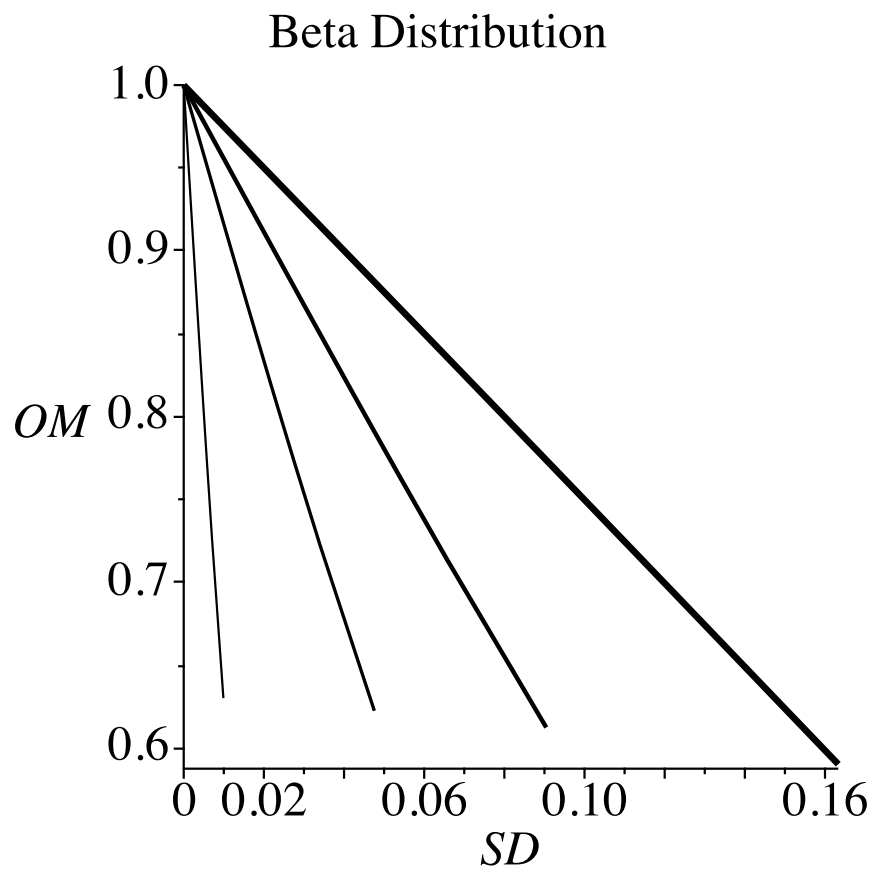

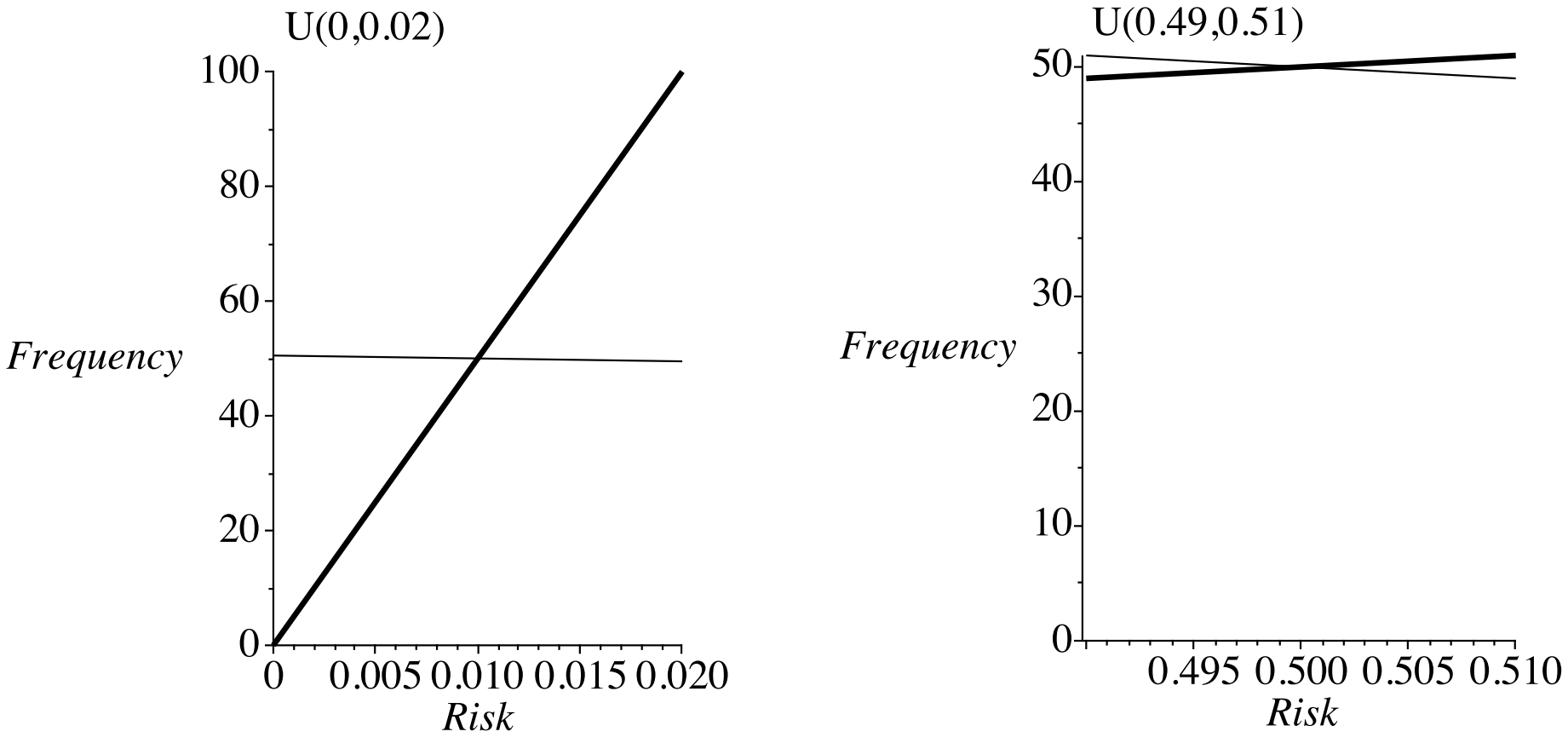

| Distribution | Probability density function, mean, and standard deviation | Area under the receiver operating characteristics curve | Overlap measure |
|---|---|---|---|
| **Uniform** | $U(r) = \frac{1}{b-a}$ | $\frac{1}{2} + \frac{(a-1)}{3(a+b-2)} - \frac{a}{3(a+b)}$ | $\frac{2a^2 + (4b-5)a + 2b^2 - 3b}{2(a+b)(a+b-2)}$ |
| | $r_{mean} = \frac{a+b}{2}$ | $\frac{1}{2} + \frac{b-a}{12\, r_{mean}(1-r_{mean})}$ | $1 - \frac{(b-a)}{8\, r_{mean}(1-r_{mean})}$ |
| | $SD = \frac{b-a}{2\sqrt{3}}$ | $\frac{1}{2} + \frac{\sqrt{3}}{6\, r_{mean}(1-r_{mean})} * SD$ | $1 - \left(\frac{\sqrt{3}}{4\, r_{mean}(1-r_{mean})}\right) * SD$ |
| **Half-sine** | $S(r) = \frac{\pi \sin\left(\frac{(r-a)\pi}{b-a}\right)}{2b-2a}$ | $\frac{1}{2} + \frac{(a-1)}{4(a+b-2)} - \frac{a}{4(a+b)}$ | $\frac{(a^2 + (2b-3)a + b^2 - b)\pi + 2a - 2b}{(a+b)\pi(a+b-2)}$ |
| | $r_{mean} = \frac{a+b}{2}$ | $\frac{1}{2} + \frac{b-a}{16\, r_{mean}(1-r_{mean})}$ | $1 - \frac{(\pi-2)(b-a)}{4\,\pi\, r_{mean}(1-r_{mean})}$ |
| | $SD = \frac{(b-a)}{2\pi}\sqrt{(\pi^2-8)}$ | $\frac{1}{2} + \frac{\pi}{8\sqrt{\pi^2-8}\, r_{mean}(1-r_{mean})} * SD$ | $1 - \left(\frac{\pi-2}{2\sqrt{\pi^2-8}\, r_{mean}(1-r_{mean})}\right) * SD$ |
| **Lognormal** | $LN(r,m,s) = \frac{1}{r\sqrt{2\pi s^2}} e^{-\frac{(\ln r - m)^2}{2s^2}}$ | | |
| | $r_{mean} = e^{m+\frac{s^2}{2}}$ | | |
| | $LN(r, r_{mean}, s) = \frac{1}{r\sqrt{2\pi s^2}} e^{-\frac{\left(\ln r - \left(\ln(r_{mean}) - \frac{s^2}{2}\right)\right)^2}{2s^2}}$ | | |
| **Beta** | $B(r) = \frac{r^{\alpha-1}(1-r)^{\beta-1}\Gamma(\alpha+\beta)}{\Gamma(\alpha)\Gamma(\beta)}$ | | |
| | $r_{mean} = \frac{\alpha}{\alpha+\beta}$ | | |
| | $SD = \sqrt{\frac{\alpha\beta}{(\alpha+\beta)^2(\alpha+\beta+1)}}$ | | |

Table of formulas